\documentclass[aps,prl,reprint,groupedaddress]{revtex4-1}
\usepackage{graphicx}
\usepackage{amsmath}

\begin{document}

\title{Learning selectivity and invariance through spatiotemporal Hebbian plasticity in a hierarchical neural network}

\author{Minjoon Kouh}
\email[]{mkouh@drew.edu}
\affiliation{Physics Department, Drew University, Madison, NJ 07940}

\begin{abstract}
When an object moves smoothly across a field of view, the identify of the object is unchanged, but the activation pattern of the photoreceptors on the retina changes drastically. One of the major computational roles of our visual system is to manage selectivity for different objects and tolerance to such identity-preserving transformations as translations or rotations. This study demonstrates that a hierarchical neural network, whose synaptic connectivities are learned competitively with Hebbian plasticity operating within a local spatiotemporal pooling range, is capable of gradually achieving feature selectivity and transformation tolerance, so that the top level neurons carry higher mutual information about object categories than a single-level neural network. Furthermore, when genetic algorithm is applied to search for a network architecture that maximizes transformation-invariant object recognition performance, in conjunction with the associative learning algorithm, it is found that deep networks outperform shallower ones.
\end{abstract}


\maketitle

\section{Introduction}
Object recognition can be considered as a task of determining whether a set of conditions that make up an ``object-ness'' is satisfied or not. From the lock-and-key mechanism of molecular receptors, to a computer algorithm for face identification, recognition process involves: selecting useful features, learning and memorizing the presence or absence of those features in different object categories, and then classifying an object in question. All recognition systems are also subject to energy, time, and other practical constraints through which their effectiveness and robustness (e.g., quickly and correctly identifying a prey or a predator) are tested. 

It is widely acknowledged that the primate visual cortex is a highly efficient recognition system, whose general level of performance is not yet surpassed by engineered systems, as we have not yet ``solved'' the recognition problem \citep{DiCarloZoccolanRust12}. Object recognition is a computationally difficult task, because it requires for the recognition system to be simultaneously selective for the differences between objects and tolerant to the changes when the same object undergoes different transformations. For example, two different sets of photoreceptors on retina would be activated when the same face appears on the left and the right sides of the visual field, while an overlapping set of photoreceptors are activated when different faces appear on the same side of the visual field. Hence, the visual cortex is charged with producing the same output even when the inputs are very different, as in the former case (i.e., tolerance). At the same time, it must generate different outputs even when the inputs are similar, as in the latter case (i.e., selectivity). A similar mathematical operation would be modulo function, where $\textrm{mod}_2(2) = \textrm{mod}_2(4) \neq \textrm{mod}_2(3)$, even though 2 is closer to 3 than to 4.

As demonstrated by a number of biologically-plausible models of visual object recognition \citep{Fukushima83, LeCun98, RiesenhuberPoggio99b, KouhPoggio08}, a hierarchical processing is one possible computational strategy to satisfy these two requirements for selectivity and tolerance. In these models, the neural population at the upper level becomes more tolerant to object transformations, by combining the responses from less tolerant neural population at the lower levels. At the same time, the neural population becomes selective for different visual features by competitively learning to transmit as much information as possible about various stimuli \citep{BellSejnowski97, OlshausenField97, Foldiak90, Stone96}.

Hebbian plasticity, which postulates that correlated activities facilitate the strengthening of neural connectivity \citep{Hebb59, DanPoo04}, is a plausible learning mechanism for selectivity and tolerance. As all objects in the world follow the laws of physics, there are predictable structure and inherent coherence in their appearances, which can be exploited by the Hebbian, associative plasticity to acquire neural response properties that are useful for object recognition \citep{Hopfield82, Foldiak90, BellSejnowski97}. Transformation tolerance can be achieved by learning temporal correlations, as an object tends to undergo its transformations smoothly in time \citep{Foldiak91, WiskottSejnowski02, Stone96}. 

Neuroanatomy and physiological studies of the primate cortex suggest that the visual information is processed hierarchically, so that the receptive fields of the visual neurons in the higher cortical areas are larger, their tuning properties are more complex, and their responses are more tolerant to translation, rotation, and scaling of a stimulus \citep{DiCarloZoccolanRust12}. A potential rationale for such a cortical organization is that constructing a highly selective yet simultaneously tolerant feature detector is computationally difficult, because a highly specific description for an object (e.g., a person with a red jacket) would not be effective in identifying the same object under different circumstances (e.g., the same person with a blue jacket). It has been observed that there is indeed a trade-off between selectivity and tolerance properties \citep{ZoccolanKouh07, SharpeeKouhReynold13}, and the hierarchical organization seems to be a strategy employed by the neural systems to deal with such an inherent trade-off. Although the trade-off of selectivity and tolerance is inevitable for a given level in the hierarchy, it is possible to gradually increase the capacity for generalization after several stages of neural computations. 

This study corroborates these intuitive ideas, and demonstrates that unsupervised learning of selectivity and tolerance can be performed with a canonical algorithm, where the same plasticity rule is applied repeatedly within a hierarchy. Other recent ``deep-learning'' approaches \citep{LeCun98, Hinton06, KouhPoggio08, GeorgeHawkins09, LeNg12} have also explored similar one-algorithm principle for configuring a recognition system, which is a theoretically attractive hypothesis for its simplicity and an evolutionarily practical solution for building a robust recognition system. Contribution of this study is in the integration of spatial and temporal learning, where temporal correlations in the stimuli endow transformation tolerance \citep{Foldiak91, Stone96, WiskottSejnowski02, StringerPerryRolls06} and spatial correlations endow selectivity to diverse visual features.

\section{Algorithm}
\begin{figure}
\begin{center}
\includegraphics[width=10cm,keepaspectratio]{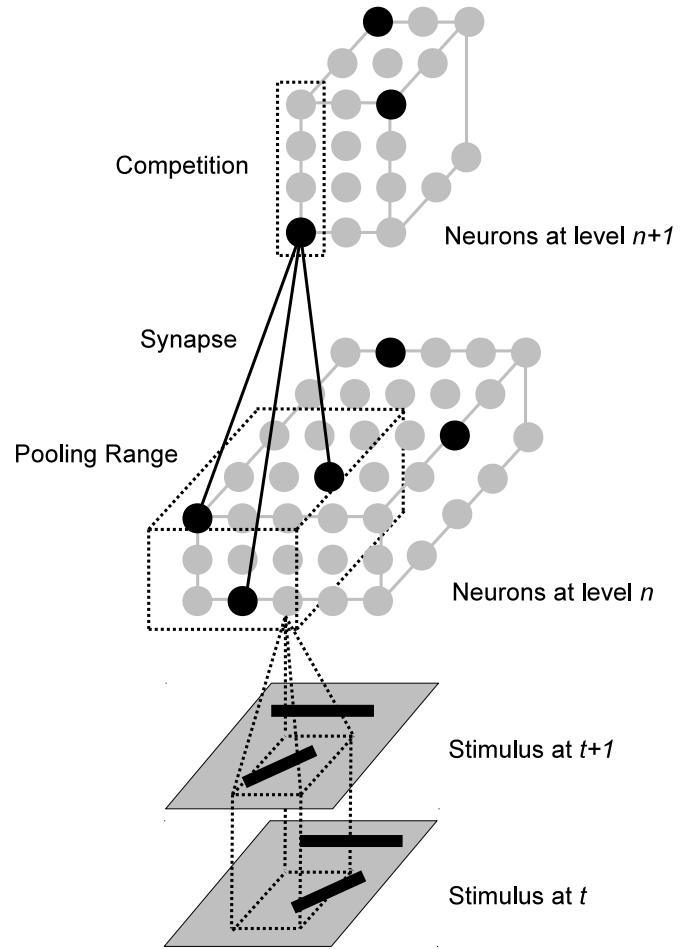}
\end{center}
\caption{\label{model} Schematic diagram of the hierarchical neural network model. Each postsynaptic neuron has a limited spatial and temporal pooling range of presynaptic neurons. There is a competition among the neurons with the same pooling range, and the winning neuron forms the synapses with the active presynaptic neurons within the pooling range. Inactive neurons are depicted with gray circles, and active ones are black. The same rule of synaptic plasticity operates throughout all the levels in the hierarchy, so that the neurons in the upper levels have larger and more complex receptive fields with a bigger range of tolerance to object transformations. Only two levels ($n$ and $n+1$) are shown here.}
\end{figure}

There are three major components in the learning algorithm: Hebbian plasticity, competition, and hierarchical architecture. The schematic diagram of the underlying neural network model is shown in Fig.~\ref{model}.

\begin{enumerate}
\item The Hebbian rule takes advantage of the consistency and persistency of object features in space and time. A synaptic connection is created, if a presynaptic neuron and a postsynaptic neuron are co-active within a predetermined spatial or temporal afferent pooling ranges. 

\item Competition allows for the neural network to acquire diverse response properties. During the learning process, a single neuron among a set of neurons with the same afferent pooling range is selected to undergo Hebbian plasticity. The other neurons are suppressed, until a different neuron is selected. The competition is based on the afferent activities. As a result, different neurons learn to specialize in representing diverse patterns of afferent activities. Once the learning phase is complete, competition is turned off. 

\item Above two algorithmic elements are embedded hierarchically, so that Hebbian learning of selectivity and tolerance starts at the lowest level and builds up sequentially and gradually to the higher levels. 
\end{enumerate}

In the implementation of the algorithm, binary neural responses were considered. The activity of a postsynaptic neuron $y_i$ is determined by a simple thresholding operation after summing its afferent activities $x_j$, which mimics the firing of an action potential when the cumulative influence of the presynaptic input brings the membrane potential above a spiking threshold. 
\begin{equation}
y_{i} = 
\begin{cases} 
1, & \mbox{if } \sum\limits_{j} w_{ij} x_{j} > \theta \\
0, & \mbox{otherwise.}
\end{cases}
\end{equation}
The threshold $\theta$ determines the sparsity of response \citep{OlshausenField97}, and in this study, it was fixed at a constant value of 0.3. A more sophisticated model of neural activity could involve a microcircuit with divisive normalization for gaussian-like selectivity and maximum-like invariance operations \citep{KouhPoggio08}.

Synaptic weights were also binary, so that a pair of presynaptic and postsynaptic neurons could be either connected or not. The synapses are initialized randomly. During the learning phase, a synapse is created between a presynaptic neuron $x_j$ and a winning postsynaptic neuron $y_i$. The winning neuron is chosen as the one with the largest presynaptic activity among a pool of neurons having the same spatial and temporal receptive field or belonging to the same cortical column \citep{Mountcastle57, Mountcastle03, GeorgeHawkins09}, even if its input has reached the threshold $\theta$. When there are multiple neurons with the same presynaptic activity, one is randomly chosen. For a winning $y_i$,
\begin{equation}
w_{ij} = 
\begin{cases}
1, & \mbox{if } x_{j} = 1, \mbox{where } j \in \mathcal{P} \\
0, & \mbox{otherwise.}
\end{cases}
\end{equation}
The pooling range, $\mathcal{P}$, is a local spatiotemporal window from which presynaptic responses can be pooled. Such local pooling is crucial for gradually building up selectivity and tolerance, as well as for creating a well-connected network with minimal amount of wiring. Local temporal pooling is consistent with the idea of trace learning \citep{Foldiak91, Stone96} and spike-timing-dependent plasticity (STDP) \citep{DanPoo04, MasquelierThorpe07}. Two different types of pooling ranges are implemented as shown in Table \ref{param}. One type has a spatially large, temporally short range, so that the postsynaptic neurons develop representation for different spatial features. The other type has a spatially small, temporally long range, so that the temporal sequence of an object transformation is learned. These two types are interleaved hierarchically, so that the neural population at the lowest level learns spatially, and the next level, temporally, and the next level, spatially, and so forth, in a similar fashion as the alternating layers of simple and complex cells in the feedforward, hierarchical models of visual cortex \citep{KouhPoggio08, RiesenhuberPoggio99b, Fukushima83}.

Another important implementation detail is a mechanism to mark the beginning of each learning cycle, when a new round of competition begins. The onset of an eye movement or phases of various brain waves may serve this role of resetting the learning cycle, while each eye fixation, lasting a few hundred milliseconds on average, is long enough to capture an object going through a short sequence of motion or other transformations. 

\begin{table}[!t]
{\label{param}Simulation parameters}
{\begin{tabular}{r r r}\hline
Level & Pooling Range & Number of Neurons \\
\hline
0 & -                        & 36 $\times$ 36 $\times$ 1 \\
1 & 6 $\times$ 6 $\times$ 1  & 16 $\times$ 16 $\times$ 8 \\
2 & 2 $\times$ 2 $\times$ 4  &  8 $\times$  8 $\times$ 4 \\
3 & 6 $\times$ 6 $\times$ 1  &  2 $\times$  2 $\times$ 8 \\
4 & 2 $\times$ 2 $\times$ 4  &  1 $\times$  1 $\times$ 8 \\
\hline 
\end{tabular}}
\caption{The zeroth level is for the stimulus image, whose dimension is 36$\times$36 pixels. 2500 (500 exemplars at 5 different positions) images were used for learning, and 1000 (200 new exemplars) images were used for calculating the mutual information in Fig.~\ref{info}. The pooling parameters are listed as horizontal $\times$ vertical $\times$ temporal ranges. The number of neurons in each level is determined by the original stimulus size, pooling range, and the degree of overlap between the neighboring pooling ranges. The last number indicates the number of neurons having the same pooling range.}
\end{table}

\section{Results}

\subsection{Learning from simple stimuli}

\begin{figure}
\begin{flushleft}
(a) Sample stimuli\\
\includegraphics[scale=0.6]{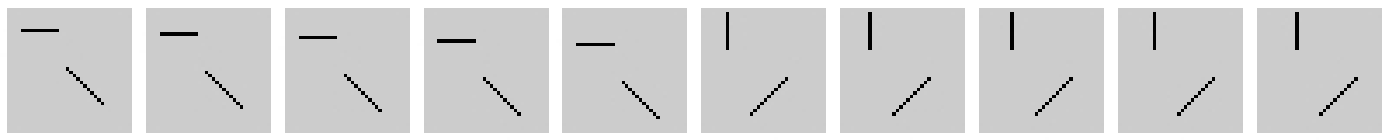} \\
\vspace{2mm}
(b) Synaptic connectivity of 8 neurons from level 1 \\
\includegraphics[scale=0.6]{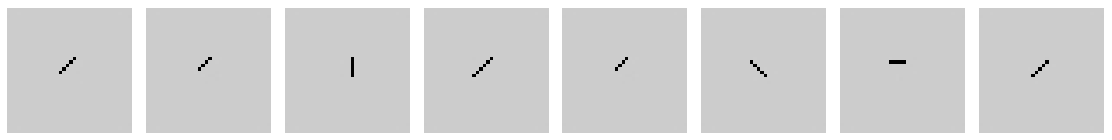} \\
\vspace{2mm}
(c) Synaptic connectivity of 2 neurons from level 2 \\
\includegraphics[scale=0.6]{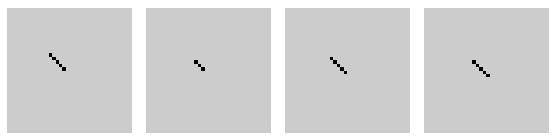} \\
\vspace{2mm}
\includegraphics[scale=0.6]{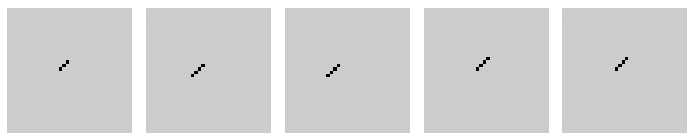} \\
\vspace{2mm}
(d) Synaptic connectivity of 2 neurons from level 3 \\
\includegraphics[scale=0.6]{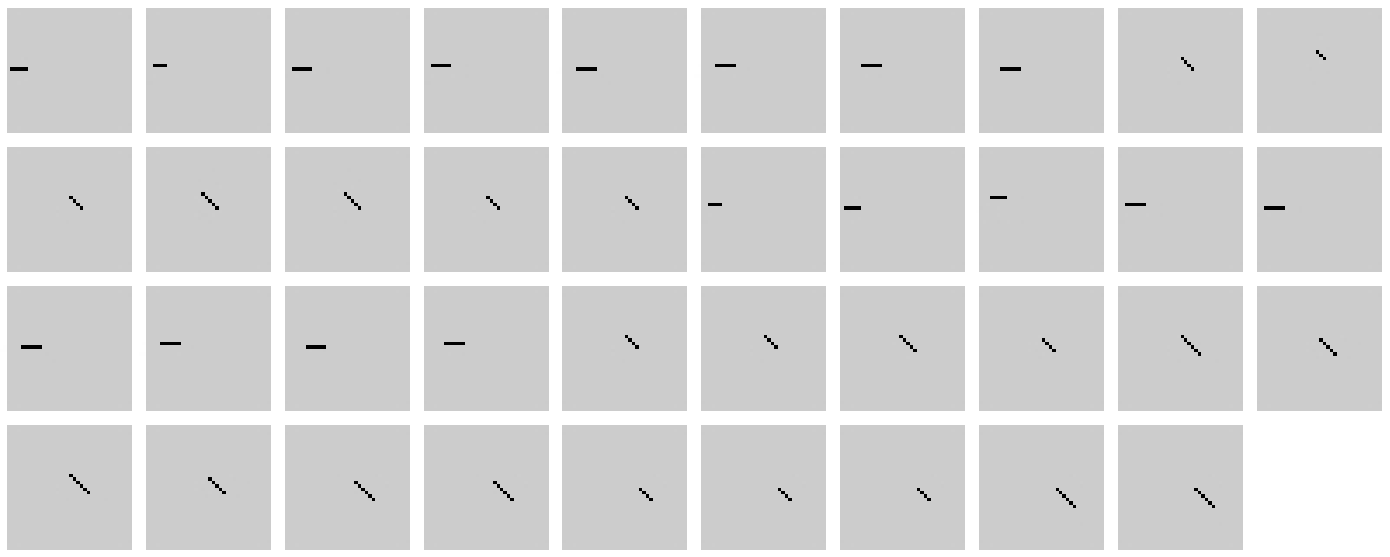} \\
\vspace{2mm}
\includegraphics[scale=0.6]{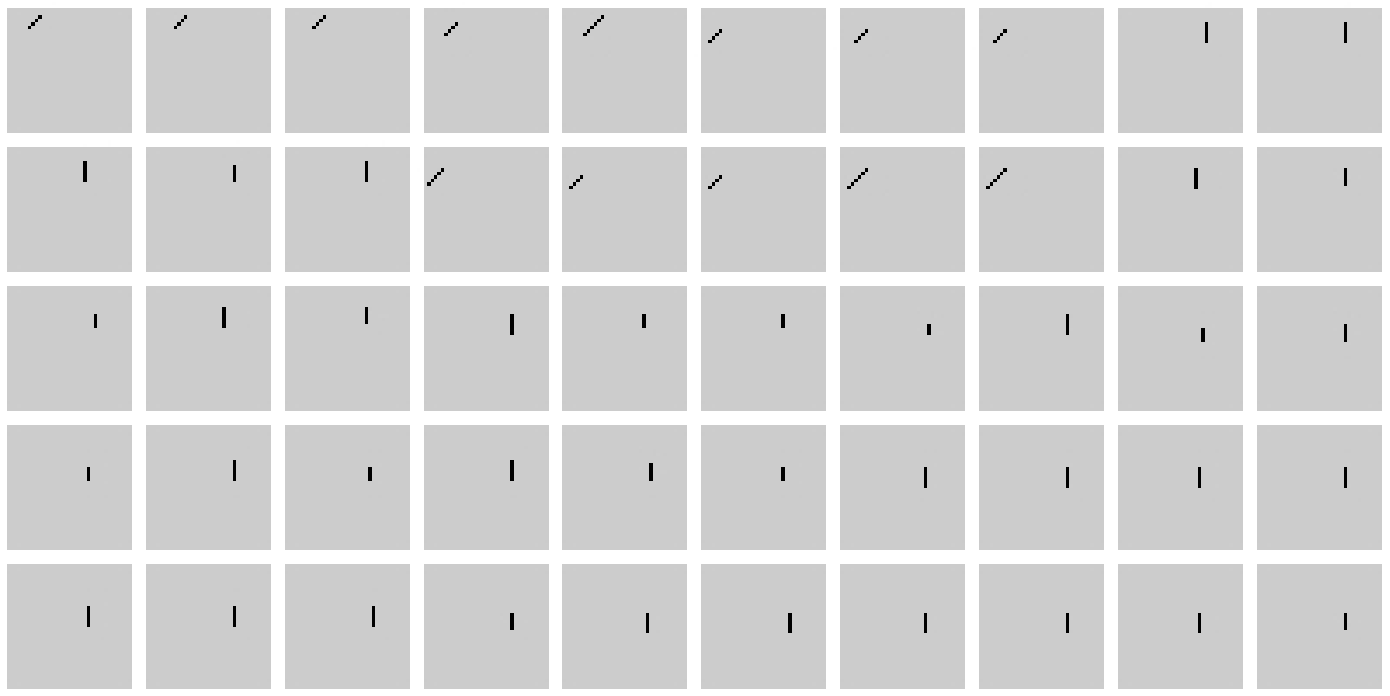} \\
\end{flushleft}
\caption{\label{w} The spatiotemporal Hebbian learning achieves selective and invariant representation. (a) Ten frames of sample stimulus images. The first five frames show the first object, composed of a specific, co-occurring pair of oriented lines, and the next five frames show the second object. Both objects undergo smooth translations. Each pixel can be considered as the activity of the photoreceptors on the retina. (b) Each panel corresponds to the learned connectivity with the photoreceptors for each neuron in level 1. Hence, each neuron becomes active when the stimulus matches the pattern of synaptic connectivity. (c) The first sample neuron connects with four presynaptic neurons with the same orientation selectivity at different positions. The second sample neuron connects with five presynaptic neurons. These two neurons in level 2 are orientation-selective and translation-tolerant. (d) The first sample neuron pools from the neurons in level 2, which in turn pool from 39 neurons in level 1 overall. This neuron shows the selectivity for one of the objects (i.e., a combination of ``---'' and ``$\backslash$'') and has a large receptive field. The second sample neuron is selective for the other object. The neurons in the next level (not shown) pool from even more presynaptic neuron, while maintaining consistent object selectivity.}
\end{figure}

The basic operation of the learning algorithm is demonstrated with images of two toy ``objects,'' where each object is composed of a specific, co-occurring pair of simpler shape elements, namely line segments of four different orientations. The first object was composed of ``---'' and ``$\backslash$'', and the second, ``$\vert$'' and ``/''. Each shape element appeared at non-overlapping random positions, and was translated smoothly in random directions, as shown in Fig.~\ref{w}(a). As a simplistic analogy, consider two different people: one of them always wearing a blue hat and a green shirt, and the other wearing orange pants and red shoes. A visual recognition system could learn to distinguish these two people by forming a representation about the four different clothing items and their combinations.

When the hierarchical neural network with the proposed learning algorithm is presented with multiple exemplars of these two objects, the lowest level neurons, having the smallest receptive fields, learn the fragments of the four shape elements, as shown in Fig.~\ref{w}(b). Since these shape elements appear at different positions, the neurons in the next level pool temporally and develop synaptic connectivity with the presynaptic neurons that have the same orientation selectivity at different positions, therefore acquiring tolerance to translation, as shown in Fig.~\ref{w}(c). This is the same two-level feedforward model of simple and complex cells in the primary visual cortex, constructed through Hebbian plasticity \citep{HubelWiesel68}. The neurons in the upper levels with larger receptive fields learn to associate the co-occurring shape elements and form a representation for the two objects, as shown in Fig.~\ref{w}(d). Such connectivity is also consistent with the bimodal orientation tuning of the intermediate cortical areas, such as V2 and V4 \citep{Gallant96, DavidGallant06, AnzaiVanEssen07, HegdeVanEssen07}. Furthermore, associative object representation between two subsequent primate temporal areas may arise from the same hierarchical learning \citep{HirabayashiMiyashita13}.

The trend of gradual and systematic increase in selectivity and tolerance can also be analyzed with mutual information between neural response $r$ and stimulus $s$, $I(r,s)$, which captures the amount of uncertainty about $s$ removed by knowing $r$ \citep{StringerPerryRolls06}.
\begin{equation}
I(r,s) = \sum\limits_{r,s} P(r,s)\log_2{\frac{P(r,s)}{P(r)P(s)}}.
\end{equation}
Since individual shape elements could appear at various positions, the response of the neurons in the lower levels, which are not yet tolerant to translation nor selective for the specific combinations of line segments, would yield low information, while the response of the neurons in the higher levels would yield larger information about the stimulus category. For example, if the response of a neuron is $r = \{0, 0, 1, 1\}$, when two exemplars of one object $O_1$ is shown, followed by two exemplars of different object $O_2$ (i.e., $s = \{O_1, O_1, O_2, O_2\}$), the mutual information is 1 bit, since an umambiguous object recognition can be performed with the neural response $r$. If the neural responses to the same sequence of stimuli were $r = \{1, 0, 1, 0\}$, the mutual information between the stimuli and the response would be zero.

\begin{figure}
\begin{center}
\includegraphics[scale=0.8]{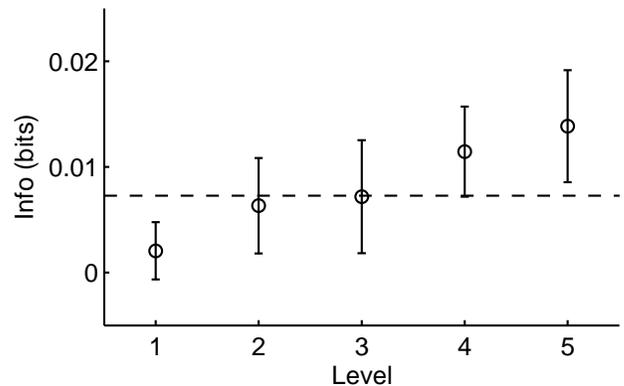}
\end{center}
\caption{\label{info} Mutual information between neural response and stimuli. After synaptic connectivities have been learned from a set of training images, response to a new set images was used to evaluate mutual information. The dotted horizontal line indicates the mutual information of a single-level neural network model, learned with linear regression.}
\end{figure}

As shown in Fig.~\ref{info} and as expected from Fig.~\ref{w}, the responses of the neural population in the higher levels on average carry more information about the objects. A similar trend is found when object classification performance was compared, using the neural responses from a higher (e.g., inferotemporal cortex) and lower (e.g., V4) visual cortical areas \citep{RustDiCarlo10}.

The most important advantage of such a hierarchical construction of selectivity and tolerance is its generalization capacity. When a single-level neural network is constructed with linear regression and a threshold operation using the same training images as before, its mutual information to the same test images is lower than the average mutual information of the hierarchical model, as shown in Fig.~\ref{info}, even when the weights of the single-level model were allowed to be non-binary real values. The threshold was chosen so that the probability of response was equal to the average response probability of the top level neurons in the hierarchical model, which was about 0.01. In other words, a sparsity-matched, single-level neural network carries less information than a multi-level neural network, because the inputs are nonlinearly transformed.

\subsection{Learning from natural images and optimizing the architecture with genetic algorithm}
Going beyond learning from simple stimuli, a set of gray-scale natural images, collected from the internet, have been applied to the proposed learning algorithm. Each image was presented over several time steps during which it was translated along a randomly chosen direction, simulating a temporally coherent motion. After learning from 250 images, object recognition performance of the learned neural network was measured with the NIST database of handwritten digits. For biological plausibility and simplicity, classification was based on the nearest neighbor scheme of how close the population response vector from the top level was to the average vectors of 10 different categories (digits 0, 1, ..., 9). 

The nearest neighbor classifier was trained on 500 images of centered handwritten digits, and the classification performance was measured with 100 test images that were different from the training images and translated by 3 pixels (corresponding to approximately 10 percent of the dimension of the digits, whose sizes were 28x28 pixels) from the center. Then, a genetic algorithm \citep{Anastasio} was used to search for an optimal network architecture, using the test performance as a fitness function. This approach was motivated by the idea that the global network architecture may be optimized over a longer, evoluationary time scale, while local features are learned over a shorter time scale through visual experiences. For example, the degree of expression of neural growth factors could determine the size of dentdritic arbor \citep{McAllisterKatz95} and spatiotemporal pooling ranges, which would dtermine the number of hierarchical layers necessary to cover a desired field of view. Furthermore, the recognition performance on the tranformed test images would be a more relevant fitness function for the goal of building a selective and invariant visual system.

\begin{figure}
\begin{center}
\includegraphics[scale=0.8]{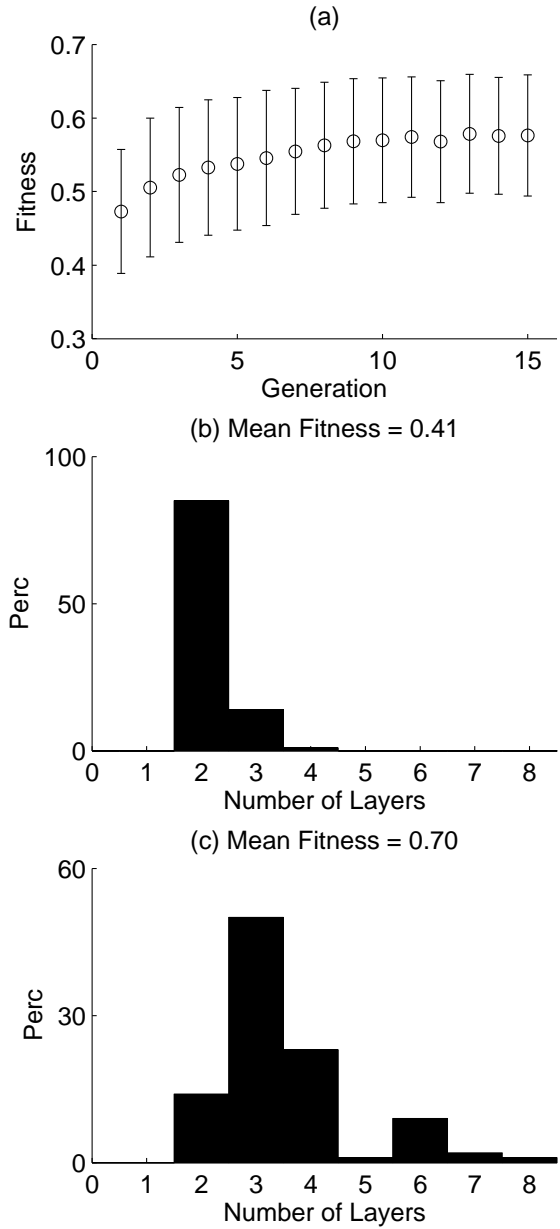}
\end{center}
\caption{\label{ga} Genetic algorithm for optimizing the network architecture. (a) The fitness of neural networks, as defined by their classification performance on translated hand-written digit images (chance = 0.1), increases over generations. The genetic algorithm finds that deeper networks on average perform better than the shallower ones, as shown by the histograms of the bottom (b) and the top (c) 100 networks. Optimized parameters were the spatial and temporal pooling ranges, as well as the number of neurons that share the same receptive field.}
\end{figure}

In this particular implementation of genetic algorithm, six parameters in the proposed learning algorithm, have been expressed as a set of binary numbers, which would go through crossover and point mutations. At each generation, two most fittest members wereee mated to produce 8 offsprings. The six parameters that were explored were: spatial and temporal pooling ranges plus the number of neurons in a competition pool for two adjacent layers. In other words, six fixed parameters in Table \ref{param} (namely 6 and 1 for the pooling range and 8 for the number of neurons in odd levels and 2, 4, and 4 in even levels) are now allowed to vary at each generation of the genetic algorithm. Fig.~\ref{ga} shows that on average the neural networks in the later generations perform bettern than their ancestors. More importantly, when the number of hierarhical layers in the neural networks are compared, it is found that deeper networks outperform shallower ones in recognizing translated objects.

\section{Discussion}

This study has demonstrated that the proposed algorithm is capable of learning selectivity and tolerance by associating spatially or temporally co-occurring features. It follows the idea that the visual system takes advantage of the physical coherence of objects in the world \citep{Foldiak91, Stone96, WiskottSejnowski02, StringerPerryRolls06} and the hierarchical, local pooling is an important architectural principle of the neural system \citep{DiCarloZoccolanRust12, KouhPoggio08}. As a proof of concept, this study has explored simple toy stimuli only, so testing the scalability of the algorithm with natural images to various transformation sequences \citep{StringerPerryRolls06,LeNg12}, as well as investigating the applicability of the same canonical algorithm in other sensory modalities \citep{Kanan13}, will be important.



\bibliography{biblio_all}

\begin{thebibliography}{33}%
\makeatletter
\providecommand \@ifxundefined [1]{%
 \@ifx{#1\undefined}
}%
\providecommand \@ifnum [1]{%
 \ifnum #1\expandafter \@firstoftwo
 \else \expandafter \@secondoftwo
 \fi
}%
\providecommand \@ifx [1]{%
 \ifx #1\expandafter \@firstoftwo
 \else \expandafter \@secondoftwo
 \fi
}%
\providecommand \natexlab [1]{#1}%
\providecommand \enquote  [1]{``#1''}%
\providecommand \bibnamefont  [1]{#1}%
\providecommand \bibfnamefont [1]{#1}%
\providecommand \citenamefont [1]{#1}%
\providecommand \href@noop [0]{\@secondoftwo}%
\providecommand \href [0]{\begingroup \@sanitize@url \@href}%
\providecommand \@href[1]{\@@startlink{#1}\@@href}%
\providecommand \@@href[1]{\endgroup#1\@@endlink}%
\providecommand \@sanitize@url [0]{\catcode `\\12\catcode `\$12\catcode
  `\&12\catcode `\#12\catcode `\^12\catcode `\_12\catcode `\%12\relax}%
\providecommand \@@startlink[1]{}%
\providecommand \@@endlink[0]{}%
\providecommand \url  [0]{\begingroup\@sanitize@url \@url }%
\providecommand \@url [1]{\endgroup\@href {#1}{\urlprefix }}%
\providecommand \urlprefix  [0]{URL }%
\providecommand \Eprint [0]{\href }%
\providecommand \doibase [0]{http://dx.doi.org/}%
\providecommand \selectlanguage [0]{\@gobble}%
\providecommand \bibinfo  [0]{\@secondoftwo}%
\providecommand \bibfield  [0]{\@secondoftwo}%
\providecommand \translation [1]{[#1]}%
\providecommand \BibitemOpen [0]{}%
\providecommand \bibitemStop [0]{}%
\providecommand \bibitemNoStop [0]{.\EOS\space}%
\providecommand \EOS [0]{\spacefactor3000\relax}%
\providecommand \BibitemShut  [1]{\csname bibitem#1\endcsname}%
\let\auto@bib@innerbib\@empty
\bibitem [{\citenamefont {{DiCarlo}}\ \emph {et~al.}(2012)\citenamefont
  {{DiCarlo}}, \citenamefont {Zoccolan},\ and\ \citenamefont
  {Rust}}]{DiCarloZoccolanRust12}%
  \BibitemOpen
  \bibfield  {author} {\bibinfo {author} {\bibfnamefont {J.~J.}\ \bibnamefont
  {{DiCarlo}}}, \bibinfo {author} {\bibfnamefont {D.}~\bibnamefont {Zoccolan}},
  \ and\ \bibinfo {author} {\bibfnamefont {N.~C.}\ \bibnamefont {Rust}},\
  }\href@noop {} {\bibfield  {journal} {\bibinfo  {journal} {Neuron}\ }\textbf
  {\bibinfo {volume} {73}},\ \bibinfo {pages} {415} (\bibinfo {year}
  {2012})}\BibitemShut {NoStop}%
\bibitem [{\citenamefont {Fukushima}\ \emph {et~al.}(1983)\citenamefont
  {Fukushima}, \citenamefont {Miyake},\ and\ \citenamefont
  {Ito}}]{Fukushima83}%
  \BibitemOpen
  \bibfield  {author} {\bibinfo {author} {\bibfnamefont {K.}~\bibnamefont
  {Fukushima}}, \bibinfo {author} {\bibfnamefont {S.}~\bibnamefont {Miyake}}, \
  and\ \bibinfo {author} {\bibfnamefont {T.}~\bibnamefont {Ito}},\ }\href@noop
  {} {\bibfield  {journal} {\bibinfo  {journal} {IEEE Transactions on Systems,
  Man and Cybernetics}\ }\textbf {\bibinfo {volume} {13}},\ \bibinfo {pages}
  {826} (\bibinfo {year} {1983})}\BibitemShut {NoStop}%
\bibitem [{\citenamefont {{LeCun}}\ \emph {et~al.}(1998)\citenamefont
  {{LeCun}}, \citenamefont {Bottou}, \citenamefont {Bengio},\ and\
  \citenamefont {Haffner}}]{LeCun98}%
  \BibitemOpen
  \bibfield  {author} {\bibinfo {author} {\bibfnamefont {Y.}~\bibnamefont
  {{LeCun}}}, \bibinfo {author} {\bibfnamefont {L.}~\bibnamefont {Bottou}},
  \bibinfo {author} {\bibfnamefont {Y.}~\bibnamefont {Bengio}}, \ and\ \bibinfo
  {author} {\bibfnamefont {P.}~\bibnamefont {Haffner}},\ }\href@noop {}
  {\bibfield  {journal} {\bibinfo  {journal} {Proceedings of the IEEE}\
  }\textbf {\bibinfo {volume} {86}},\ \bibinfo {pages} {2278} (\bibinfo {year}
  {1998})}\BibitemShut {NoStop}%
\bibitem [{\citenamefont {Riesenhuber}\ and\ \citenamefont
  {Poggio}(1999)}]{RiesenhuberPoggio99b}%
  \BibitemOpen
  \bibfield  {author} {\bibinfo {author} {\bibfnamefont {M.}~\bibnamefont
  {Riesenhuber}}\ and\ \bibinfo {author} {\bibfnamefont {T.}~\bibnamefont
  {Poggio}},\ }\href@noop {} {\bibfield  {journal} {\bibinfo  {journal} {Nature
  Neuroscience}\ }\textbf {\bibinfo {volume} {2}},\ \bibinfo {pages} {1019}
  (\bibinfo {year} {1999})}\BibitemShut {NoStop}%
\bibitem [{\citenamefont {Kouh}\ and\ \citenamefont
  {Poggio}(2008)}]{KouhPoggio08}%
  \BibitemOpen
  \bibfield  {author} {\bibinfo {author} {\bibfnamefont {M.}~\bibnamefont
  {Kouh}}\ and\ \bibinfo {author} {\bibfnamefont {T.}~\bibnamefont {Poggio}},\
  }\href@noop {} {\bibfield  {journal} {\bibinfo  {journal} {Neural
  Computation}\ }\textbf {\bibinfo {volume} {20}},\ \bibinfo {pages} {1427}
  (\bibinfo {year} {2008})}\BibitemShut {NoStop}%
\bibitem [{\citenamefont {Bell}\ and\ \citenamefont
  {Sejnowski}(1997)}]{BellSejnowski97}%
  \BibitemOpen
  \bibfield  {author} {\bibinfo {author} {\bibfnamefont {A.~J.}\ \bibnamefont
  {Bell}}\ and\ \bibinfo {author} {\bibfnamefont {T.~J.}\ \bibnamefont
  {Sejnowski}},\ }\href@noop {} {\bibfield  {journal} {\bibinfo  {journal}
  {Vision Res.}\ }\textbf {\bibinfo {volume} {23}},\ \bibinfo {pages} {3327}
  (\bibinfo {year} {1997})}\BibitemShut {NoStop}%
\bibitem [{\citenamefont {Olshausen}\ and\ \citenamefont
  {Field}(1997)}]{OlshausenField97}%
  \BibitemOpen
  \bibfield  {author} {\bibinfo {author} {\bibfnamefont {B.~A.}\ \bibnamefont
  {Olshausen}}\ and\ \bibinfo {author} {\bibfnamefont {D.~J.}\ \bibnamefont
  {Field}},\ }\href@noop {} {\bibfield  {journal} {\bibinfo  {journal} {Vision
  Research}\ }\textbf {\bibinfo {volume} {37}},\ \bibinfo {pages} {3311}
  (\bibinfo {year} {1997})}\BibitemShut {NoStop}%
\bibitem [{\citenamefont {F\"{o}ldi\'{a}k}(1990)}]{Foldiak90}%
  \BibitemOpen
  \bibfield  {author} {\bibinfo {author} {\bibfnamefont {P.}~\bibnamefont
  {F\"{o}ldi\'{a}k}},\ }\href@noop {} {\bibfield  {journal} {\bibinfo
  {journal} {Biological Cybernetics}\ }\textbf {\bibinfo {volume} {64}},\
  \bibinfo {pages} {165} (\bibinfo {year} {1990})}\BibitemShut {NoStop}%
\bibitem [{\citenamefont {Stone}(1996)}]{Stone96}%
  \BibitemOpen
  \bibfield  {author} {\bibinfo {author} {\bibfnamefont {J.~V.}\ \bibnamefont
  {Stone}},\ }\href@noop {} {\bibfield  {journal} {\bibinfo  {journal} {Neural
  Computation}\ }\textbf {\bibinfo {volume} {8}},\ \bibinfo {pages} {1463}
  (\bibinfo {year} {1996})}\BibitemShut {NoStop}%
\bibitem [{\citenamefont {Hebb}(1959)}]{Hebb59}%
  \BibitemOpen
  \bibfield  {author} {\bibinfo {author} {\bibfnamefont {D.~O.}\ \bibnamefont
  {Hebb}},\ }\href@noop {} {\bibfield  {journal} {\bibinfo  {journal} {Brain}\
  }\textbf {\bibinfo {volume} {82}},\ \bibinfo {pages} {260} (\bibinfo {year}
  {1959})}\BibitemShut {NoStop}%
\bibitem [{\citenamefont {Dan}\ and\ \citenamefont {Poo}(2004)}]{DanPoo04}%
  \BibitemOpen
  \bibfield  {author} {\bibinfo {author} {\bibfnamefont {Y.}~\bibnamefont
  {Dan}}\ and\ \bibinfo {author} {\bibfnamefont {M.~M.}\ \bibnamefont {Poo}},\
  }\href@noop {} {\bibfield  {journal} {\bibinfo  {journal} {Neuron}\ }\textbf
  {\bibinfo {volume} {44}},\ \bibinfo {pages} {23} (\bibinfo {year}
  {2004})}\BibitemShut {NoStop}%
\bibitem [{\citenamefont {Hopfield}(1982)}]{Hopfield82}%
  \BibitemOpen
  \bibfield  {author} {\bibinfo {author} {\bibfnamefont {J.~J.}\ \bibnamefont
  {Hopfield}},\ }\href@noop {} {\bibfield  {journal} {\bibinfo  {journal}
  {Proc.~Natl.~Acad.~Sci.~USA}\ }\textbf {\bibinfo {volume} {79}},\ \bibinfo
  {pages} {2554} (\bibinfo {year} {1982})}\BibitemShut {NoStop}%
\bibitem [{\citenamefont {F\"{o}ldi\'{a}k}(1991)}]{Foldiak91}%
  \BibitemOpen
  \bibfield  {author} {\bibinfo {author} {\bibfnamefont {P.}~\bibnamefont
  {F\"{o}ldi\'{a}k}},\ }\href@noop {} {\bibfield  {journal} {\bibinfo
  {journal} {Neural Computation}\ }\textbf {\bibinfo {volume} {3}},\ \bibinfo
  {pages} {194} (\bibinfo {year} {1991})}\BibitemShut {NoStop}%
\bibitem [{\citenamefont {Wiskott}\ and\ \citenamefont
  {Sejnowski}(2002)}]{WiskottSejnowski02}%
  \BibitemOpen
  \bibfield  {author} {\bibinfo {author} {\bibfnamefont {L.}~\bibnamefont
  {Wiskott}}\ and\ \bibinfo {author} {\bibfnamefont {T.~J.}\ \bibnamefont
  {Sejnowski}},\ }\href@noop {} {\bibfield  {journal} {\bibinfo  {journal}
  {Neural Computation}\ }\textbf {\bibinfo {volume} {14}},\ \bibinfo {pages}
  {715} (\bibinfo {year} {2002})}\BibitemShut {NoStop}%
\bibitem [{\citenamefont {Zoccolan}\ \emph {et~al.}(2007)\citenamefont
  {Zoccolan}, \citenamefont {Kouh}, \citenamefont {Poggio},\ and\ \citenamefont
  {{DiCarlo}}}]{ZoccolanKouh07}%
  \BibitemOpen
  \bibfield  {author} {\bibinfo {author} {\bibfnamefont {D.}~\bibnamefont
  {Zoccolan}}, \bibinfo {author} {\bibfnamefont {M.}~\bibnamefont {Kouh}},
  \bibinfo {author} {\bibfnamefont {T.}~\bibnamefont {Poggio}}, \ and\ \bibinfo
  {author} {\bibfnamefont {J.~J.}\ \bibnamefont {{DiCarlo}}},\ }\href@noop {}
  {\bibfield  {journal} {\bibinfo  {journal} {Journal of Neuroscience}\
  }\textbf {\bibinfo {volume} {27}},\ \bibinfo {pages} {12292} (\bibinfo {year}
  {2007})}\BibitemShut {NoStop}%
\bibitem [{\citenamefont {Sharpee}\ \emph {et~al.}(2013)\citenamefont
  {Sharpee}, \citenamefont {Kouh},\ and\ \citenamefont
  {Reynolds}}]{SharpeeKouhReynold13}%
  \BibitemOpen
  \bibfield  {author} {\bibinfo {author} {\bibfnamefont {T.~O.}\ \bibnamefont
  {Sharpee}}, \bibinfo {author} {\bibfnamefont {M.}~\bibnamefont {Kouh}}, \
  and\ \bibinfo {author} {\bibfnamefont {J.~H.}\ \bibnamefont {Reynolds}},\
  }\href@noop {} {\bibfield  {journal} {\bibinfo  {journal}
  {Proc.~Natl.~Acad.~Sci.~USA}\ }\textbf {\bibinfo {volume} {110}},\ \bibinfo
  {pages} {11618} (\bibinfo {year} {2013})}\BibitemShut {NoStop}%
\bibitem [{\citenamefont {Hinton}\ and\ \citenamefont
  {Salakhutdinov}(2006)}]{Hinton06}%
  \BibitemOpen
  \bibfield  {author} {\bibinfo {author} {\bibfnamefont {G.~E.}\ \bibnamefont
  {Hinton}}\ and\ \bibinfo {author} {\bibfnamefont {R.~R.}\ \bibnamefont
  {Salakhutdinov}},\ }\href@noop {} {\bibfield  {journal} {\bibinfo  {journal}
  {Science}\ }\textbf {\bibinfo {volume} {313}},\ \bibinfo {pages} {504}
  (\bibinfo {year} {2006})}\BibitemShut {NoStop}%
\bibitem [{\citenamefont {George}\ and\ \citenamefont
  {Hawkins}(2009)}]{GeorgeHawkins09}%
  \BibitemOpen
  \bibfield  {author} {\bibinfo {author} {\bibfnamefont {D.}~\bibnamefont
  {George}}\ and\ \bibinfo {author} {\bibfnamefont {J.}~\bibnamefont
  {Hawkins}},\ }\href@noop {} {\bibfield  {journal} {\bibinfo  {journal}
  {{PLoS} Computational Biology}\ }\textbf {\bibinfo {volume} {5}},\ \bibinfo
  {pages} {e1000532} (\bibinfo {year} {2009})}\BibitemShut {NoStop}%
\bibitem [{\citenamefont {Le}\ \emph {et~al.}(2011)\citenamefont {Le},
  \citenamefont {Ranzato}, \citenamefont {Monga}, \citenamefont {Devin},
  \citenamefont {Chen}, \citenamefont {Corrado}, \citenamefont {Dean},\ and\
  \citenamefont {Ng}}]{LeNg12}%
  \BibitemOpen
  \bibfield  {author} {\bibinfo {author} {\bibfnamefont {Q.~V.}\ \bibnamefont
  {Le}}, \bibinfo {author} {\bibfnamefont {M.}~\bibnamefont {Ranzato}},
  \bibinfo {author} {\bibfnamefont {R.}~\bibnamefont {Monga}}, \bibinfo
  {author} {\bibfnamefont {M.}~\bibnamefont {Devin}}, \bibinfo {author}
  {\bibfnamefont {K.}~\bibnamefont {Chen}}, \bibinfo {author} {\bibfnamefont
  {G.~S.}\ \bibnamefont {Corrado}}, \bibinfo {author} {\bibfnamefont
  {J.}~\bibnamefont {Dean}}, \ and\ \bibinfo {author} {\bibfnamefont {A.~Y.}\
  \bibnamefont {Ng}},\ }\href@noop {} {\bibfield  {journal} {\bibinfo
  {journal} {Proceedings of the International Conference on Machine Learning}\
  } (\bibinfo {year} {2011})}\BibitemShut {NoStop}%
\bibitem [{\citenamefont {Stringer}\ \emph {et~al.}(2006)\citenamefont
  {Stringer}, \citenamefont {Perry}, \citenamefont {Rolls},\ and\ \citenamefont
  {Proske}}]{StringerPerryRolls06}%
  \BibitemOpen
  \bibfield  {author} {\bibinfo {author} {\bibfnamefont {S.~M.}\ \bibnamefont
  {Stringer}}, \bibinfo {author} {\bibfnamefont {G.}~\bibnamefont {Perry}},
  \bibinfo {author} {\bibfnamefont {E.~T.}\ \bibnamefont {Rolls}}, \ and\
  \bibinfo {author} {\bibfnamefont {J.}~\bibnamefont {Proske}},\ }\href@noop {}
  {\bibfield  {journal} {\bibinfo  {journal} {Biological cybernetics}\ }\textbf
  {\bibinfo {volume} {94}},\ \bibinfo {pages} {128} (\bibinfo {year}
  {2006})}\BibitemShut {NoStop}%
\bibitem [{\citenamefont {Mountcastle}(1957)}]{Mountcastle57}%
  \BibitemOpen
  \bibfield  {author} {\bibinfo {author} {\bibfnamefont {V.~B.}\ \bibnamefont
  {Mountcastle}},\ }\href@noop {} {\bibfield  {journal} {\bibinfo  {journal}
  {Journal of Neurophysiology}\ }\textbf {\bibinfo {volume} {20}},\ \bibinfo
  {pages} {408} (\bibinfo {year} {1957})}\BibitemShut {NoStop}%
\bibitem [{\citenamefont {Mountcastle}(2003)}]{Mountcastle03}%
  \BibitemOpen
  \bibfield  {author} {\bibinfo {author} {\bibfnamefont {V.~B.}\ \bibnamefont
  {Mountcastle}},\ }\href@noop {} {\bibfield  {journal} {\bibinfo  {journal}
  {Cereb.~Cortex}\ }\textbf {\bibinfo {volume} {13}},\ \bibinfo {pages} {2}
  (\bibinfo {year} {2003})}\BibitemShut {NoStop}%
\bibitem [{\citenamefont {Masquelier}\ and\ \citenamefont
  {Thorpe}(2007)}]{MasquelierThorpe07}%
  \BibitemOpen
  \bibfield  {author} {\bibinfo {author} {\bibfnamefont {T.}~\bibnamefont
  {Masquelier}}\ and\ \bibinfo {author} {\bibfnamefont {S.~J.}\ \bibnamefont
  {Thorpe}},\ }\href@noop {} {\bibfield  {journal} {\bibinfo  {journal} {{PLoS}
  Computational Biology}\ }\textbf {\bibinfo {volume} {3}},\ \bibinfo {pages}
  {e31} (\bibinfo {year} {2007})}\BibitemShut {NoStop}%
\bibitem [{\citenamefont {Hubel}\ and\ \citenamefont
  {Wiesel}(1968)}]{HubelWiesel68}%
  \BibitemOpen
  \bibfield  {author} {\bibinfo {author} {\bibfnamefont {D.}~\bibnamefont
  {Hubel}}\ and\ \bibinfo {author} {\bibfnamefont {T.}~\bibnamefont {Wiesel}},\
  }\href@noop {} {\bibfield  {journal} {\bibinfo  {journal} {Journal of
  Physiology}\ }\textbf {\bibinfo {volume} {195}},\ \bibinfo {pages} {215}
  (\bibinfo {year} {1968})}\BibitemShut {NoStop}%
\bibitem [{\citenamefont {Gallant}\ \emph {et~al.}(1996)\citenamefont
  {Gallant}, \citenamefont {Connor}, \citenamefont {Rakshit}, \citenamefont
  {Lewis},\ and\ \citenamefont {Van~Essen}}]{Gallant96}%
  \BibitemOpen
  \bibfield  {author} {\bibinfo {author} {\bibfnamefont {J.~L.}\ \bibnamefont
  {Gallant}}, \bibinfo {author} {\bibfnamefont {C.~E.}\ \bibnamefont {Connor}},
  \bibinfo {author} {\bibfnamefont {S.}~\bibnamefont {Rakshit}}, \bibinfo
  {author} {\bibfnamefont {J.~W.}\ \bibnamefont {Lewis}}, \ and\ \bibinfo
  {author} {\bibfnamefont {D.~C.}\ \bibnamefont {Van~Essen}},\ }\href@noop {}
  {\bibfield  {journal} {\bibinfo  {journal} {Journal of Neurophysiology}\
  }\textbf {\bibinfo {volume} {76}},\ \bibinfo {pages} {2718} (\bibinfo {year}
  {1996})}\BibitemShut {NoStop}%
\bibitem [{\citenamefont {David}\ \emph {et~al.}(2006)\citenamefont {David},
  \citenamefont {Hayden},\ and\ \citenamefont {Gallant}}]{DavidGallant06}%
  \BibitemOpen
  \bibfield  {author} {\bibinfo {author} {\bibfnamefont {S.~V.}\ \bibnamefont
  {David}}, \bibinfo {author} {\bibfnamefont {B.~Y.}\ \bibnamefont {Hayden}}, \
  and\ \bibinfo {author} {\bibfnamefont {J.}~\bibnamefont {Gallant}},\
  }\href@noop {} {\bibfield  {journal} {\bibinfo  {journal} {Journal of
  Neurophysiology}\ }\textbf {\bibinfo {volume} {96}},\ \bibinfo {pages} {3492}
  (\bibinfo {year} {2006})}\BibitemShut {NoStop}%
\bibitem [{\citenamefont {Anzai}\ \emph {et~al.}(2007)\citenamefont {Anzai},
  \citenamefont {Peng},\ and\ \citenamefont {Van~Essen}}]{AnzaiVanEssen07}%
  \BibitemOpen
  \bibfield  {author} {\bibinfo {author} {\bibfnamefont {A.}~\bibnamefont
  {Anzai}}, \bibinfo {author} {\bibfnamefont {X.}~\bibnamefont {Peng}}, \ and\
  \bibinfo {author} {\bibfnamefont {D.~C.}\ \bibnamefont {Van~Essen}},\
  }\href@noop {} {\bibfield  {journal} {\bibinfo  {journal} {Nature
  neuroscience}\ }\textbf {\bibinfo {volume} {10}},\ \bibinfo {pages} {1313}
  (\bibinfo {year} {2007})}\BibitemShut {NoStop}%
\bibitem [{\citenamefont {Hegde}\ and\ \citenamefont
  {Van~Essen}(2007)}]{HegdeVanEssen07}%
  \BibitemOpen
  \bibfield  {author} {\bibinfo {author} {\bibfnamefont {J.}~\bibnamefont
  {Hegde}}\ and\ \bibinfo {author} {\bibfnamefont {D.~C.}\ \bibnamefont
  {Van~Essen}},\ }\href@noop {} {\bibfield  {journal} {\bibinfo  {journal}
  {Cereb.~Cortex}\ }\textbf {\bibinfo {volume} {17}},\ \bibinfo {pages} {1100}
  (\bibinfo {year} {2007})}\BibitemShut {NoStop}%
\bibitem [{\citenamefont {Hirabayashi}\ \emph {et~al.}(2013)\citenamefont
  {Hirabayashi}, \citenamefont {Takeuchi}, \citenamefont {Tamura},\ and\
  \citenamefont {Miyashita}}]{HirabayashiMiyashita13}%
  \BibitemOpen
  \bibfield  {author} {\bibinfo {author} {\bibfnamefont {T.}~\bibnamefont
  {Hirabayashi}}, \bibinfo {author} {\bibfnamefont {D.}~\bibnamefont
  {Takeuchi}}, \bibinfo {author} {\bibfnamefont {K.}~\bibnamefont {Tamura}}, \
  and\ \bibinfo {author} {\bibfnamefont {Y.}~\bibnamefont {Miyashita}},\
  }\href@noop {} {\bibfield  {journal} {\bibinfo  {journal} {Science}\ }\textbf
  {\bibinfo {volume} {341}},\ \bibinfo {pages} {191} (\bibinfo {year}
  {2013})}\BibitemShut {NoStop}%
\bibitem [{\citenamefont {Rust}\ and\ \citenamefont
  {{DiCarlo}}(2010)}]{RustDiCarlo10}%
  \BibitemOpen
  \bibfield  {author} {\bibinfo {author} {\bibfnamefont {N.~C.}\ \bibnamefont
  {Rust}}\ and\ \bibinfo {author} {\bibfnamefont {J.~J.}\ \bibnamefont
  {{DiCarlo}}},\ }\href@noop {} {\bibfield  {journal} {\bibinfo  {journal}
  {Journal of Neuroscience}\ }\textbf {\bibinfo {volume} {30}},\ \bibinfo
  {pages} {12978} (\bibinfo {year} {2010})}\BibitemShut {NoStop}%
\bibitem [{\citenamefont {Anastasio}(2010)}]{Anastasio}%
  \BibitemOpen
  \bibfield  {author} {\bibinfo {author} {\bibfnamefont {T.~J.}\ \bibnamefont
  {Anastasio}},\ }\href@noop {} {\emph {\bibinfo {title} {Tutorial on neural
  systems modeling}}}\ (\bibinfo  {publisher} {Sinauer},\ \bibinfo {year}
  {2010})\BibitemShut {NoStop}%
\bibitem [{\citenamefont {McAllister}\ \emph {et~al.}(1995)\citenamefont
  {McAllister}, \citenamefont {Lo},\ and\ \citenamefont
  {Katz}}]{McAllisterKatz95}%
  \BibitemOpen
  \bibfield  {author} {\bibinfo {author} {\bibfnamefont {A.~K.}\ \bibnamefont
  {McAllister}}, \bibinfo {author} {\bibfnamefont {D.~C.}\ \bibnamefont {Lo}},
  \ and\ \bibinfo {author} {\bibfnamefont {L.~C.}\ \bibnamefont {Katz}},\
  }\href@noop {} {\bibfield  {journal} {\bibinfo  {journal} {Neuron}\ }\textbf
  {\bibinfo {volume} {15}},\ \bibinfo {pages} {791} (\bibinfo {year}
  {1995})}\BibitemShut {NoStop}%
\bibitem [{\citenamefont {Kanan}(2013)}]{Kanan13}%
  \BibitemOpen
  \bibfield  {author} {\bibinfo {author} {\bibfnamefont {C.}~\bibnamefont
  {Kanan}},\ }\href@noop {} {\bibfield  {journal} {\bibinfo  {journal} {{PLoS}
  One}\ }\textbf {\bibinfo {volume} {8}},\ \bibinfo {pages} {e54088} (\bibinfo
  {year} {2013})}\BibitemShut {NoStop}%
\end{thebibliography}%

\end{document}